\documentclass[a4paper,fleqn,usenatbib]{mnras}
\newcommand{\iac}{intermediate-age clusters}
\newcommand{\dinbas}{DynBaS}
\newcommand{\dinbasuno}{DynBaS1D}
\newcommand{\dinbasdos}{DynBaS2D}
\newcommand{\dinbastres}{DynBaS3D}
\newcommand{\msun}{M$_{\odot}$}

\usepackage{txfonts}
\usepackage[T1]{fontenc}
\usepackage{ae,aecompl}

\usepackage{graphicx}	% Including figure files
\usepackage{amssymb}	% Extra maths symbols

\title[Is the $v_{esc}$ in Star Clusters Linked to Extended SFH?]{Is the Escape Velocity in Star Clusters Linked to Extended
Star Formation Histories?  Using NGC 7252:\,W3 as a Test Case}
\author[I. Cabrera-Ziri et al.]{I. Cabrera-Ziri$^{1,2}$\thanks{ICZ: icabrera@eso.org},   N. Bastian$^{2}$, M. Hilker$^{1}$, B. Davies$^{2}$, F. Schweizer$^3$,\linebreak \newauthor J. M. D. Kruijssen$^{4,5}$, A. Mej\'ia-Narv\'aez$^6$, F. Niederhofer$^{7,8}$, T. D. Brandt$^9$\thanks{NASA Sagan Fellow},\linebreak \newauthor M. Rejkuba$^{1,7}$, G. Bruzual$^{10}$, G. Magris$^5$
\\
$^{1}$ European Southern Observatory, Karl-Schwarzschild-Stra{\ss}e 2, D-85748 Garching bei M{\"u}nchen, Germany\\
$^{2}$ Astrophysics Research Institute, Liverpool John Moores University, 146 Brownlow Hill, Liverpool L3 5RF, UK\\
$^3$ Carnegie Observatories, 813 Santa Barbara Street, Pasadena, CA 91101, USA\\
$^4$ Max-Planck Institut f{\"u}r Astrophysik, Karl-Schwarzschild-Stra{\ss}e 1,  D-85748 Garching bei M{\"u}nchen, Germany\\
$^5$ Astronomisches Rechen-Institut, Zentrum f\"{u}r Astronomie der Universit\"{a}t Heidelberg, M\"{o}nchhofstra\ss e 12-14, 69120 Heidelberg, Germany\\
$^6$ Centro de Investigaciones de Astronom\'ia, A.P. 264, M\'erida 5101-A, Venezuela\\
$^7$ Excellence Cluster Origin and Structure of the Universe, Boltzmannstr. 2, D-85748 Garching bei M{\"u}nchen, Germany\\
$^8$ Universit{\"a}ts-Sternwarte M{\"u}nchen, Scheinerstra{\ss}e 1, D-81679 M{\"u}nchen, Germany\\
$^9$ School of Natural Sciences, Institute for Advanced Study, Princeton, NJ, USA\\
$^{10}$ Instituto de Radioastronom\'ia y Astrof\'isica, IRyA, UNAM, Campus Morelia, A.P. 3-72, C.P. 58089 Michoac\'an, Mexico
}

\date{Accepted 2015 December 21.  Received 2015 November 17; in original form 2015 July 23}

\pubyear{2015}

\begin{document}
\label{firstpage}
\pagerange{\pageref{firstpage}--\pageref{lastpage}}
\maketitle

\begin{abstract}
The colour-magnitude diagrams of some \iac\ (1--2 Gyr) star clusters show unexpectedly broad main-sequence turnoffs, raising the possibility that these clusters have experienced more than one episode of star formation.
%\textbf{Since the discovery that the main-sequence turnoff in some intermediate-age (1--2 Gyr) clusters is broader than what is expected for a simple stellar population,} the possibility that these clusters have experienced multiple star formation episodes has been considered in order to explain its origin.
%some authors have considered the possibility that these clusters have experienced multiple star formation episodes in order to explain its origin. 
Such a scenario predicts the existence of an extended main sequence turn off (eMSTO)
%This line of thought suggests that this phenomenon would be observed
 only in clusters with escape velocities above a certain threshold ($>15$ km s$^{-1}$), which would allow them to retain or accrete gas that eventually would fuel a secondary extended star-formation episode.
%  If true, the intermediate-age clusters observed with an eMSTO must have undergone an extended star-formation event of a few hundred ($\sim400$) Myr at an early age that contributed an important fraction of the current cluster mass.
   This paper presents a test of this scenario based on the study of the young and massive cluster NGC 7252:\,W3. We use the \emph{HST} photometry from WFPC2 and WFC3 images obtained with %$F336W,~F439W,~F475W,~F555W,~F775W$ and $F814W$
   UV and optical filters, as well as MagE echellette spectrograph data from the Las Campanas Clay 6.5m telescope, in order to construct the observed UV/optical SED of NGC 7252:\,W3. The observations are then compared with synthetic spectra based on different star formation histories consistent with those of the eMSTO clusters. We find that the SED of this cluster is best fitted by a synthetic spectrum with a single stellar population of age $570^{+70}_{-62}$ Myr and mass $1.13^{+0.14}_{-0.13}\times 10^8$ \msun, confirming earlier works on NGC 7252:W3. We also estimate the lower limit on the central escape velocity of    
%   and current escape velocity larger than
    193 km s$^{-1}$.
%     \textbf{confirming earlier works on NGC 7252:\,W3}.
      We rule out extended star-formation histories, like those inferred for the eMSTO clusters in the Magellanic Clouds, at high confidence. We conclude that the escape velocity of a cluster does not dictate whether a cluster can undergo extended periods of star formation. 
\end{abstract}


\section{Introduction}
\label{sec:intro}

In the last decade, precise photometry of stars in several intermediate-age (1--2 Gyr) clusters in the Small and Large Magellanic Clouds (SMC and LMC) has revealed an extended main-sequence turnoff (eMSTO) in their optical colour--magnitude diagrams (CMDs). Many studies have interpreted these observations as extended episodes of star formation lasting hundreds of Myr (e.g. \citealt{Mackey07,Mackey:2008p2587,Goudfrooij:2011p2248,Goudfrooij:2011p2254,Milone:2009p2291,Rubele:2013p2304}). To sustain extended or multiple episodes of star formation a cluster needs to be able to retain stellar ejecta or acquire new gas, which poses limits on the escape velocities of clusters hosting eMSTO. The current escape velocities of the clusters hosting eMSTOs are modest (3--20 km s$^{-1}$, see \citealt{Goudfrooij:2011p2254}), and in many cases are below those estimated for young massive clusters that do not show evidence for extended star-formation episodes lasting $>30$~Myr \citep{Bastian:2013p2199,Longmore14,Kruijssen14}.

However, \cite{Goudfrooij:2011p2248} have suggested that the intermediate-age clusters may have lost a significant fraction of their stars since their birth, hence their initial escape velocities were much higher, potentially $>15$~km/s, which the authors claim is a limit above which clusters can retain their stellar ejecta.  In this scenario, a first generation of stars forms in a near instantaneous burst, which is then followed by a lull that lasts between a few Myr and a few hundred Myr, and then by a further Gaussian-shaped extended star formation episode that lasts a few hundred Myr.  In order to have enough matter available to form the second generation of stars a large amount of material needs to be accreted from the clusters' surroundings, because the material shed by the first generation is not sufficient to form the observed numbers of second generation stars. However, we note that a plausible mechanism for this accretion has not yet been identified.
%
%(as the material shed during the evolution of first-generation stars is insufficient to form the observed numbers of inferred second-generation stars) a large amount of gas would also be required to be accreted from the surroundings, although the mechanism behind this accretion has not been identified.

%It has also been proposed that there exists a
A link between the eMSTO phenomenon in \iac\ and the chemical anomalies found in globular clusters (GCs, see \citealt{Gratton:2012p2005}) has been proposed, suggesting a common evolution of massive clusters independent of the environment and time of formation (e.g. \citealt{Keller11,cs11,Goudfrooij:2011p2254,Goudfrooij:2011p2248,Goudfrooij:2014p2918}, hereafter G11a,b and G14). The above scenario and its link with multiple populations in GCs have been tested by searching for abundance spreads within the eMSTO clusters, which are expected to be observed if these clusters are self-enriched.
%should exist if self-enrichment is happening within the clusters. 
However, no evidence for abundance spreads have been found in the clusters with eMSTOs (\citealt{Mucciarelli:2008p2339,Mucciarelli:2014p2575}, Mackey et al.\ in prep.). Hence, self-enrichment is unlikely to have happened in the eMSTO clusters, and the eMSTO phenomenon does not appear to be linked to multiple populations in GCs.

The lack of ongoing star formation within young ($< 1$ Gyr) massive ($>10^4$ \msun) clusters (YMCs) and the lack of extended star formation histories (SFHs) in resolved YMCs in the LMC are seemingly at variance with the age spreads inferred from the eMSTOs \citep{Bastian:2013p2199,Niederhofer15}. In addition, some post-main-sequence evolutionary phases (e.g. the subgiant branch and red clump) of eMSTO clusters appear to be incompatible with the extended SFHs inferred from the analysis of their turnoffs (\citealt{BN15,Li14,Niederhofer15b}, although see \citealt{Goudfrooij15}). This may indicate that alternative explanations are needed for these phenomena such as stellar evolutionary effects, e.g. stellar rotation, as has been suggested and explored by \citet{Bastian09}, \citet{Li14} and \citet{Brandt15b}.

In two recent studies of the resolved stellar populations in the young ($\sim300$ Myr) massive ($\sim10^5$ \msun) LMC cluster NGC 1856 \citep{Milone15,Correnti15}, the authors found evidence for an eMSTO at young ages for the first time. Although these authors suggest a prolonged star-formation episode as the origin of the eMSTO, further analyses of post-main-sequence stars are necessary to see whether they are in agreement with this interpretation. However, note that the proposed duration of the secondary star-formation episode is significantly shorter than that inferred for the 1--2 Gyr old clusters.  \cite{Niederhofer15b} have shown that there is a strong relation between the age of the cluster and the inferred age spread, suggesting that stellar evolutionary effects are the cause of the eMSTO phenomenon.

It has been suggested that the intermediate-age (1--2~Gyr) cluster population found in the LMC/SMC arose after these galaxies underwent a strong starburst during this epoch due to a three-body interaction with the Milky Way (G14). However, estimates of the SFH for both galaxies do not show clear evidence for such a burst, as the star-formation rate seems to be constant during this period within a factor of $\sim2$, cf. \cite{Harris09,Weisz13}. The scenarios that propose extended star formation episodes as the origin of the eMSTO, also suggest that massive intermediate-age clusters, due to their large initial gravitational potential wells, were capable of retaining (and accreting) gas from which a second stellar generation was formed. Accordingly, one should see---in young massive clusters with escape velocities in excess of $\sim15$~km s$^{-1}$---clear signatures of younger generations of stars that formed after the initial, main burst in their integrated colours and SEDs. In this paper, we address the issue of the origin of the eMSTO of \iac\ by analysing the SED of W3, a YMC in the merger remnant NGC 7252 that has an escape velocity in excess of 193 km s$^{-1}$. 

%As discussed above, it has been suggested that massive intermediate-age clusters, due to their large initial gravitational potential wells, were capable of retaining (and accreting) gas from which a second stellar generation was formed. Accordingly, one should see---in young massive clusters with escape velocities in excess of $\sim15$~km s$^{-1}$---clear signatures of younger generations of stars that formed after the initial, main burst in their integrated colours and SEDs. In this paper, we address the issue of the origin of the eMSTO of \iac\ by analysing the SED of W3, a YMC in the merger remnant NGC 7252 that has an escape velocity in excess of 180 km s$^{-1}$. While this cluster formed in a galaxy merger, an environment quite different from that of the LMC/SMC, we note that it has been suggested that the reason why nearly all intermediate-age (1--2~Gyr) clusters in the LMC/SMC host eMSTOs is that these galaxies underwent a strong starburst during this epoch due to a three-body interaction with the Milky Way (G14).  However, estimates of the SFH for both galaxies do not show clear evidence for such a burst, as the star-formation rate seems to be constant during this period within a factor of $\sim2$, cf. \cite{Harris09,Weisz13}.

The cluster NGC 7252:\,W3 is an excellent candidate to test these scenarios given that with a mass of $\sim 10^8$ \msun\ \citep{SS98,Maraston04} it is the most massive young cluster known to date. The age of W3 is constrained by numerical simulations of NGC 7252, which suggest that its last major merger event took place about 600 Myr ago \citep{Hibbard:1995p2917,Chien:2010p2916}. This event has been proposed to trigger the star formation episode that gave birth to the YMC population observed in NGC 7252  \citep{Whitmore93,Miller97,SS98}. The age of $\sim 600$ Myr for W3, places it right in the range when the extended star-formation episode should be going on (or just have ceased), according to the SFHs inferred for eMSTO clusters by G14. All this makes this cluster ideal to test whether the eMSTO of \iac\ has its origin in extended periods of star formation.

%Additionally, there are further constraints on the cluster age, as numerical simulations suggest that the two major disk galaxies involved in the NGC 7252 merger collided about 620 Myr ago \citep{Hibbard:1995p2917,Chien:2010p2916}.  This has been proposed to have triggered the star-formation episode that gave birth to the YMC population observed in this system \citep{Whitmore93,Miller97,SS98}.

The paper is organised as follows: In \S \ref{sec:data} we present the \emph{HST} photometry and the MagE spectrum of W3. In \S \ref{sec:dinbas} we show the procedure used to estimate the age, mass and escape velocity of W3. The experiments with synthetic SEDs are described in \S \ref{sec:exp}.  Finally, we discuss our results and present our conclusions in \S \ref{sec:discussion} and \S\ref{sec:conclusions}, respectively.

\section{Data}
\label{sec:data}

We took the WFC3 photometry of W3 from \cite{Bastian13-7252} (bands $F336W,~F475W$ and $F775W$).\footnote{\emph{HST} programme GO-11554 (PI: Bastian; see \citealt{Bastian13-7252})}Additionally, we performed aperture photometry on WFPC2 images taken with the filters $F336W,~F439W$, $F555W$ and $F814W$.\footnote{\emph{HST} programme GO-5416 (PI: Whitmore; see \citealt{Miller97}).} The \emph{HST} pipeline processed images were first cleaned from cosmic rays by
%For this we first removed cosmic rays from the images
 using the {\sevensize  LACOS IRAF}\footnote{{\sevensize IRAF} is distributed by the National Optical Astronomy Observatories, which is operated by the Association of Universities for Research in Astronomy, Inc., under cooperative agreement with the National Science Foundation.} routine \citep{vanDokkum01}, and then we performed aperture photometry with the 
% then we used the
  task {\sevensize PHOT} from {\sevensize DAOPHOT} under {\sevensize IRAF} 
%  to do the photometry on these images using
  adopting the same aperture sizes as \cite{Bastian13-7252}, i.e. a circular aperture with 0.4\arcsec\ radius centred on W3 and a sky annulus of inner radius 1.325\arcsec\ with a width of 0.25\arcsec.
%0.48" with a diameter of 0.08". \fbox{i changed these values}
The results of our photometry on the WFPC2 images can be found in Table \ref{tab:phot}.

\begin{table}
 \caption{WFPC2 photometry of NGC 7252: W3.}\label{tab:phot}
\begin{center}
 \begin{tabular}{@{}c c c c@{}}
 \hline 
 $F336W$ & $F439W$ & $F555W$ & $F814W$  \\
  (mag) & (mag) & (mag) & (mag) \\
\hline
$18.64\pm 0.06$ & $18.41\pm0.02$ & $18.19\pm0.13$ & $17.36\pm0.06$  \\
\hline
\end{tabular}
\end{center}
\end{table}

A spectrum of this cluster was obtained on Aug 23rd of 2009 with
the MagE echellette spectrograph \citep{Marshall08} on the Clay 6.5-m
telescope at Las Campanas.
The $10\arcsec\!\times 0\farcs7$ slit was placed across the cluster at
parallactic angle.
The 2.3 hr total exposure was broken into seven 20 minute subexposures,
during which the airmass decreased from 1.49 to 1.04 and the seeing
was $\sim$\,$0\farcs7$.
To permit flux calibration, six standard stars were also observed at
parallactic angle throughout the night.

The reduction of the MagE spectrum included pipeline processing to
flat-field and coadd frames, rectify spectral orders, calibrate
wavelengths, and subtract the galaxy-plus-sky background spectrum.
The final extracted spectrum of W3 covers the wavelength range
3300\,--\,8250~\AA, extracted from orders 18\,--\,8, at a spectral
resolution of $R \approx 5500$.
Due to small wiggles in the spectral continuum introduced by the
digital splicing together of the various overlapping orders, we
here restrict our use of the spectrum to its highest-quality range
of 3300\,--\,5500~\AA. 
However, the overall SED of W3 is a good representation of the actual flux levels of this cluster, as we will show via the good agreement with \emph{HST} photometry in \S \ref{sec:exp} (see below).

\section{Age and mass of  NGC 7252:\,W3}
\label{sec:dinbas}

We used \dinbas, a Dynamical Basis Selection spectral fitting algorithm \citep{cz14,Magris15} to recover the SFH of W3. Basically, the \dinbas\ algorithm finds the best simple stellar population (SSP) template or the best linear combination of two or three SSPs templates to fit the target spectrum (with their relative weights/masses). Here, we outline briefly the basics of our spectral fitting of W3 and refer the reader interested in a more detailed discussion of the limitations and uncertainties of our technique to the publications cited above.

We performed three different fits to W3 data:

\begin{enumerate}
\item \textbf{\emph{HST} broad band photometry:} This was a full SED fit using the \emph{HST}'s WFPC2 and WPC3 photometry described in \S \ref{sec:data} and assuming an $A_V=0.083$ mag,\footnote{this value was derived from the $A_{F475W}=0.098$ reported in \cite{Bastian13-7252} and assuming $A_g=3.64 ~E_{(B-V)}$ from \cite{Jordan04}.} and a distance to NGC 7252 of 64 Mpc \citep{Bastian13-7252}. The results of all fits are presented in Table \ref{tab:dinbas}.\footnote{We also carried out the fit including the $K_s$ photometry from \cite{Maraston01} and find that the results are unaffected.}

\begin{table*}
 \caption{Results of \dinbas\ fitting: Ages and masses of each of the fit components, mass weighted ages, and total masses.}\label{tab:dinbas}
\begin{center}
 \begin{tabular}{@{}l c c c c c c@{}}
 \hline 
 Fit & Solution & $t_1$/M$_1$ & $t_2$/M$_2$ & $t_3$/M$_3$ & $<t>_{\mbox{\sc m}}$$^{(a)}$ & M$_{\mbox{tot}}$ \\
  & & (Myr / \%) & (Myr / \%) & (Myr / \%) & (Myr) &  (\msun) \\
\hline
\emph{HST} & \dinbasuno & 509 /  100 &  &  & 509  & $1.00\times 10^8$  \\
photometry & \dinbasdos & 453 / 4.3 & 509 / 95.7 & & 506 & $9.9\times 10^7$ \\
 & \dinbastres & 453 / 65.3 & 719 / 14.4 & 5,750 / 20.3 & 815 & $1.19 \times 10^8$ \\
\hline
MagE & \dinbasuno & 571 /  100 &  &  & 571 & $9.9\times 10^7$ \\
spectrum & \dinbasdos & 404 / 9.4 & 13,750 / 90.6 & & 9,883 & $6.10\times 10^8$ \\
 & \dinbastres & 360 / 2.6 & 404 / 6.2 & 13,750 / 91.2 & 10,057 & $6.24 \times 10^8$ \\
\hline
MagE & \dinbasuno & 571 /  $100$ &  &  & 571 & -  \\
normalized & \dinbasdos & 508 / $94.2$& 8,500 / $5.8$ &  & 619 & - \\
spectrum$^{(b)}$& \dinbastres & 508 / $93$ & 1,700 / $2.6$ & 8,500 / $4.4$ & 589 &  - \\
\hline
\end{tabular}
\end{center}
$^{(a)}$ $<t>_{\mbox{\sc m}} = 10^{ \sum_i \mu_i \mbox{log } t_i}$, with $\mu_i = \mbox{M}_i/\sum_i \mbox{M}_i $\\
$^{(b)}$ Since there was no continuum in this spectrum we are not able to retrieve masses.%, instead we show the\\ relative contributions of each component in percentage and used these values to calculate the mass weighted age.
\end{table*}

\item \textbf{MagE spectrum:} This was also a full SED fit, using this time the flux-calibrated MagE spectrum instead of the \emph{HST} photometry, and assuming the same $A_V$ and distance as above. The results of these fits are shown in Fig. \ref{MagEfit}. From this figure, we can see that there is a slight offset between the SED
and the synthetic stellar populations fitted by \dinbas. In order to determine whether these observed differences are due to the issues on the continuum calibration mentioned in \S \ref{sec:data}, we decided to carry out an additional fit after normalizing the continuum, in order to assess the robustness of our age determination.

\begin{figure}
\includegraphics[width= 84mm]{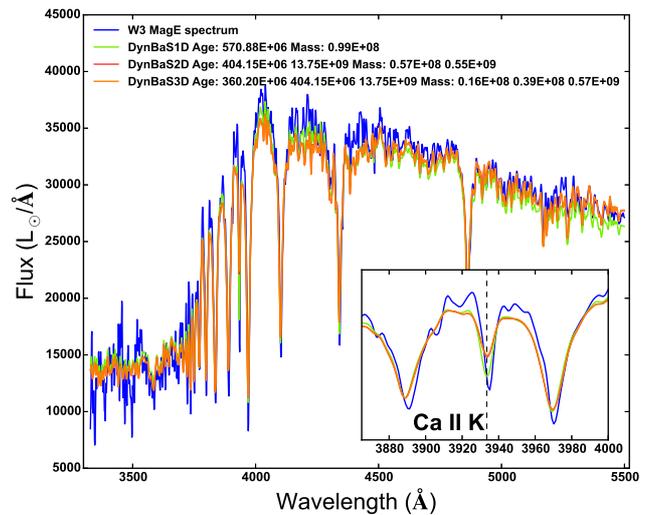}
\caption{\dinbasuno, \dinbasdos\ and \dinbastres\ fits (see the text) to the MagE spectrum of Cluster W3. The \dinbas\ fits presented here were obtained using stellar population models of solar metallicity. See text for a discussion of these results.}
\label{MagEfit}
\end{figure}

\item \textbf{MagE continuum-normalized spectrum:}  We normalized the continuum of W3's SED for our fitting. To obtain the continuum we ran a median filter of 100 \AA~width over the cluster spectrum, masking 2000 km s$^{-1}$ around the core of each Balmer line. Then we normalized the continuum for each SSP spectrum comprised in the \cite{bc03} models, using again a median filter with the same width and mask as we used for the observed spectrum. Next, we divided each model SED by its respective continuum. The spectral fit was then carried out on the normalized spectrum of W3 with the normalized SSP model spectra. The results of these fits are shown in Fig. \ref{dinbasfit}.

\begin{figure*}
\includegraphics[width= 180mm,height=84mm]{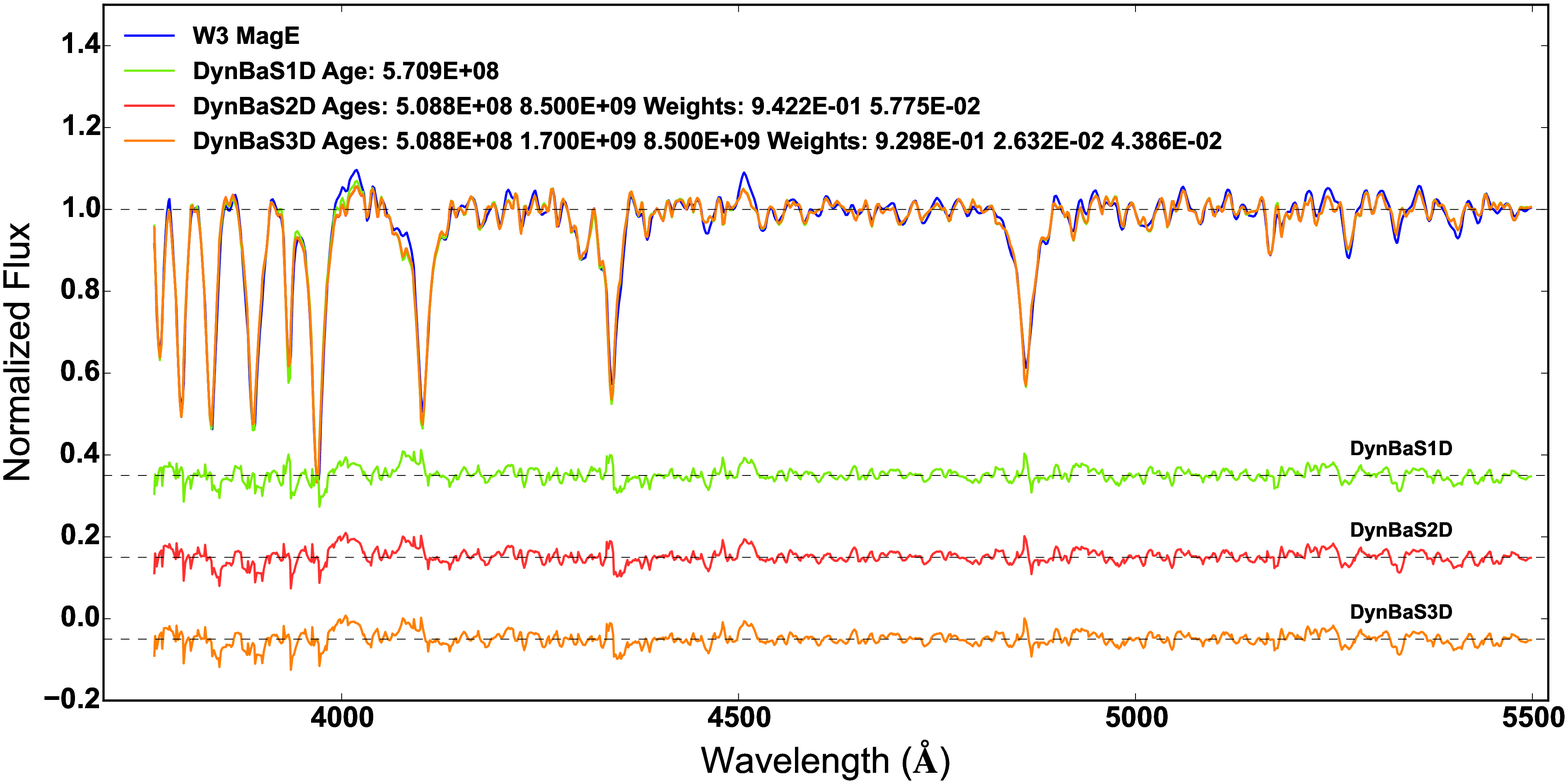}
\caption{\dinbas 1D, \dinbas 2D and \dinbas 3D fits to the continuum-normalised spectrum of W3. The \dinbas\ fits presented here were obtained using BC03 models of solar metallicity. On the bottom we plot the residuals (data - \dinbas\ fits) in the same vertical scale with 3 different offsets for clarity. The three solutions provide virtually the same spectrum. We can rule out the \dinbas 2D and \dinbas 3D solutions using the same arguments that were used for NGG 34:\,S1, see text for details.}
\label{dinbasfit}
\end{figure*}

\end{enumerate}

Table \ref{tab:dinbas} shows that all \dinbasuno\ solutions (i.e. best-fitting SSPs) are rather precise, i.e. in agreement within $\sim60$ Myr of each other. Both fits of the MagE spectrum (unnormalized and normalized) yield a cluster age of 571 Myr, while the fit to the \emph{HST} broad-band photometry yields a somewhat younger age of 509 Myr.

\begin{figure}
\includegraphics[width= 84mm]{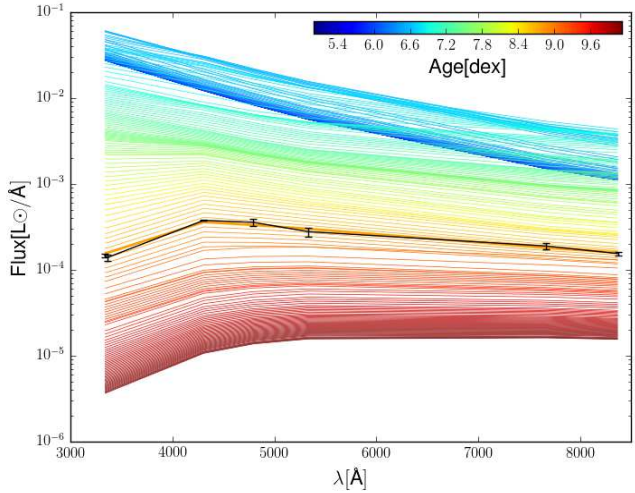}
\caption{
SED evolution of an SSP of $Z_\odot$ over the \emph{HST} filters mentioned in \S \ref{sec:data}. The thick orange line represents the SSP age determined by the \dinbas\ continuum normalized fit (570 Myr). The black line represents the \emph{HST} photometry of W3. We have scaled the stellar population templates to match the SED of W3 at 570 Myr. From this we infer a mass of $1.13\times10^8$ \msun\ for W3. All SSP templates have been attenuated to the same extinction as reported for W3.}
\label{hstfit}
\end{figure}

For both the photometry and the MagE spectrum fits, we see that the \emph{multiple-population} solutions (i.e. \dinbasdos\ and \dinbastres) consist of a young population relatively close in age to the SSP solution, i.e. about 500 Myr, and an old ($>5$ Gyr) population with a significant fraction of the total mass. The origin of these old components in the \dinbas\ fits is simple. Basically, they come from the fact that the mass-luminosity ratio increases with time for an SSP evolving passively, i.e. without any subsequent star formation events.  %Having multiple-population solutions with old and massive components that yield SEDs ``similar" to the ones of the best SSP fit is possible because the mass-luminosity ratio increases with time for an SSP of fixed mass evolving passively, i.e. with no subsequent star formation. Given that the old population is too faint with respect of the younger ones, it is possible to allocate a lot of mass in them during the fits without affecting significantly the overall SED of the younger components. --- Given that the old populations are too faint with respect of the younger ones, it is possible to allocate a considerable fraction of the mass to them without affecting the overall SED of the younger population in a significant way. 
This is shown in Fig. \ref{hstfit}, where we plot the evolution of an SSP for a fixed mass. As the figure shows, the old ($>5$ Gyr) populations are at least an order of magnitude fainter than a population of the same mass but younger age (i.e. few hundred Myr). Due to this, it is possible to allocate considerable important fractions of the mass into these old components during the fits, without affecting the overall SED of the younger populations in a significant way. In other words, if any component of the fit is attributed to an old population, due to the high mass to light ratio, it will necessarily be given a high mass.

%This makes possible that \dinbas\ multiple-population solutions with old and massive components to yield ``similar'' SEDs to the ones of the best SSP fit, as they will not affect the overall SED of the younger populations in a significant way.
However, although the overall multiple-population SED might appear to be in good agreement with the SED of the W3 and/or the SSP solution, we emphasize that the solutions with significant fractions of older components do not accurately reproduce  the Ca {\sc ii} K line, a spectral feature that is highly age sensitive (cf. Fig. \ref{MagEfit}, more on this in \S \ref{sec:discussion}).

%We conducted an experiment aiming to achieve a better reproduction of this line. Here, we carried out a \dinbas\ fit on the flux calibrated MagE spectrum, but contrary to the previous ones, we restricted the fit to very narrow spectral region (between 3700--4600 \AA). The motivation for this, is to minimize the weight of the continuum shape to the fit, as it could be driving the inclusion of older components. From this experiment, we recovered the same \dinbasuno\ solution as before (571 Myr) but this time, all the multiple population solutions were combinations of ages between 320 and 720 Myr, which yeilded mean mass-weighted ages of 570 Myr. All these mulitple-population solutions reproduced the profile of the Ca {\sc ii} K line in a much better way than the previous multiple-population solutions with old (massive) components.

As was also found for the YMC NGC~34:\,S1 in \cite{cz14}, the multiple stellar-population solutions from the continuum-normalized spectral fit require a very old and low-mass burst to have happened $\sim8.5$ Gyr ago, after which the cluster suffered a huge burst of 13--15 times the mass of the old population at a young ($\sim500$ Myr) age. These kinds of solutions would appear to be rather exotic, as discussed in \cite{cz14}. Very old populations arise when the code is forced to retrieve a multiple stellar-population solution, as they artificially improve the residuals between the observed spectrum and the stellar population templates on a very small level (i.e. less that tenths of reduced $\chi^2$).

%One of the most sensitive regions to metallicity in the optical wavelength range is between 5100--5400 \AA~given that it hosts a number of important metallicity indicators, including Mg$b$, Fe5270, Fe5335 \citep{Gonzalez93}. For our fits to the continuum-normalized spectrum of W3, we used different \cite{bc03} SSP models with metallicities of $Z = 0.4$, 1 and 2.5 $Z_\odot$ ($[Z] = -0.4$, 0 and 0.4). We found that the best fits to the spectrum (specially to the region mentioned above) were with the $Z_\odot$ templates. This is in agreement with what was reported by \citet{SS98}.

We carried out fits to the continuum-normalized spectrum of W3 using \cite{bc03} models with metallicities of $Z = 0.4$, 1 and 2.5 $Z_\odot$ ($[Z] = -0.4$, 0 and 0.4), and we found that the best fits to the spectrum (specifically, the 5100--5400 \AA~region which hosts a number of important metallicity indicators, including Mg$b$, Fe5270, Fe5335; \citealt{Gonzalez93}) were with the $Z_\odot$ templates, as also found by \citet{SS98}.

Given that the \dinbas\ SSP solution for the continuum-normalized spectrum yields the most accurate fit of all our experiments, we adopt this solution, $570$ Myr, as the age of W3. Having derived the age, we are able to infer the cluster mass, scaling the SSP of 570 Myr to the \emph{HST} photometry (corrected for extinction and distance). The match between W3 photometry and the SSP template of 570 Myr is also shown in Fig. \ref{hstfit}. Additionally, we use this figure to make a conservative estimate of the uncertainties in age/mass. We adopt as uncertainties the values of the youngest/oldest SSP ages that lie within the photometric error bars. This yields an age for W3 of $570^{+70}_{-62}$ Myr and a mass of $1.13^{+0.14}_{-0.13}\times10^8$ \msun\ (we discuss the possible degeneracies of our age determination, in Appendix \ref{sec:deg}).

These results are consistent with those from previous studies of this cluster. For example, \cite{SS98} found an age of $\sim540$ Myr and a mass of $1.8\times10^8$ \msun, a similar analysis by \cite{Maraston01} found an age of $510\pm10$ Myr while \cite{Maraston04} estimated the dynamical mass of W3 to be $8 \pm \mbox{2} \times10^7$ \msun.

\subsection{Escape velocity of  NGC 7252:\,W3}
\label{sec:exp}

We use equation (1) from \cite{Georgiev09}:

$$ v_{esc} = f_c \sqrt{\frac{M_{\rm cl}}{r_{\rm eff}}}~ (\mbox{km s}^{-1}),$$

\noindent to estimate the current escape velocity of W3. Here, $f_c$ is a coefficient which takes into account the dependence of the escape velocity on the density profile of the cluster, i.e. its concentration $c = \mbox{log}(r_t/r_c)$ where $r_t$ and $r_c$ are the tidal and core radius of the cluster, respectively.

\cite{Georgiev09} computed $f_c$ for \cite{King62} models by deprojecting the density profile and then calculating the potential as a function of radius. In order to estimate a lower limit to the escape velocity of this cluster, we adopted the smallest value for $f_c$ reported by these authors in their Table 2, i.e. 0.076. We adopt an effective radius $r_{\rm eff}=17.5~{\rm pc}$ \citep{Maraston04} and a cluster mass $M_{\rm cl}=1.13\times10^8~{\rm M}_\odot$ as determined above.
% We take $r_h$, the half-light radius, to be 17.5 pc \citep{Maraston04}. And finally we take the mass of the cluster to be $1.13\times10^8$ \msun\ as determined above. 
 With these values, we get $ v_{esc} = 193$ km s$^{-1}$ for W3, well above the $\sim15$ km s$^{-1}$ limit proposed by G11b as the threshold for any extended star-formation episode to happen (which would be responsible for the eMSTO observed in \iac). 

We emphasize that this value represents a lower limit as $f_c$ is likely higher for this cluster. According to \cite{Bastian13-7252} the tidal radius is $r_t > 500$ pc for W3. From Table 1 in G14, we derive that the mean value of $<r_c/r_{\rm eff}>=0.67$. Assuming that the core radius is $r_c\approx 0.67 \times r_{\rm eff}$, this leads to log$(r_t/r_c)= 1.63$, which would correspond to a $f_c = 0.1$, leading to a much larger $v_{esc}=254$ km s$^{-1}$ for W3.

%Assuming that the core radius is $r_c\approx 0.5 \times r_{eff}$, this leads to log$(r_t/r_c)= 1.76$, which would correspond to a $f_c = 0.103$, leading to a much larger $v_{esc}=246$ km s$^{-1}$ for W3.

\section{Synthetic SED experiments}
\label{sec:exp}

Having determined that the SFH of W3 is consistent with an SSP of 570 Myr, in this section we test how the SEDs built by using the SFHs of eMSTO clusters inferred by G14 compare to the observed SED of W3. With these experiments, we explore whether the inferred SFHs from eMSTO clusters are also compatible with the SED of this cluster, which satisfies all the characteristics of an eMSTO cluster of this age ($\sim600$ Myr), namely:

\begin{enumerate}
\item It has the age when a second episode of star formation is expected to have happened recently or be happening.
\item Currently it has an escape velocity well above the threshold suggested by G11b as being necessary to retain/accrete gas from which the second stellar generation will form.
\end{enumerate}

This makes W3 suitable to undergo, or have recently undergone, the extended star-formation episode responsible for eMSTOs according to some authors (e.g. G14 and \citealt{Correnti15}).

All the synthetic SEDs in this section and the previous one (i.e. SED fitting) were built by using \cite{bc03} stellar-population models of $Z_\odot$, assuming a \cite{Chabrier03} IMF, and computing models with `Padova 1994' evolutionary tracks \citep{Alongi93, Bressan93, Fagotto94a,Fagotto94b,Girardi96} and the stellar library STELIB \citep{LeBorgne03}. All synthetic SEDs were attenuated with the extinction value reported for W3, $A_V=0.083$ mag, by using a \cite{ccm} extinction law and $R_V=3.1$.

\subsection{Experiments with \citet{Goudfrooij:2014p2918} SFHs}

G14 analyse the eMSTOs of 18 \iac\ in the Magellanic Clouds, and report SFHs for all clusters that consist of a first, instantaneous massive burst followed by an extended period of star formation, equivalent to a small fraction of the mass of the first burst. 

We would like to emphasize that the SFHs directly inferred from the eMSTO (referred by these authors as pseudo-age distributions) have little evidence of a large initial burst. From Figs. 2, 3 and 4 of G14, we see that only an extended episode of star formation is found which can be represented reasonably well by a Gaussian distribution with a FWHM of $\sim 375$ Myr for most clusters. However, without an initial burst this would represent a problem regarding the build up of the mass of this extended star formation episode. This is because if there is no first generation, there would not be a potential well that could accrete and retain the gas from which the next generation of stars are going to be born (i.e. it would take hundreds of Myr for the clusters to build up enough mass to exceed the suggested limit of 15 km s$^{-1}$ escape velocity). In G14 they solve this problem invoking a first generation of stars which will be lost nearly entirely after the second episode of star formation takes place. By doing this they are able to build up such extended star formation episode and simultaneously match the inferred pseudo-age distributions from the eMSTO with no signs of an older and massive population. 

%The SFHs inferred by G14 from the eMSTOs of 18 \iac\ in the Magellanic Clouds consist of a first, instantaneous massive burst followed by an extended period of star formation, equivalent to a small fraction of the mass of the first burst.  From Figs.\ 2, 3 and 4 of G14, we see that the second, \emph{extended} episode of star formation can be represented accurately by a Gaussian distribution with a FWHM of $\sim 375$ Myr for most clusters.

For the present experiments, we have built the SEDs of synthetic clusters according to the same kind of star-formation episodes, i.e. a first, instantaneous burst of star formation at 570 Myr (i.e. the age recovered by \dinbas\ for W3) contributing most of the mass of the cluster, followed by an extended, Gaussian-shaped star-formation episode with a fraction of the mass of the first burst.  We have made a couple of conservative assumptions and choices:

\begin{enumerate}
\item We do not allow the final star-formation episode to extend over the last 100 Myr of the cluster life.  Young populations with current ages $<100$ Myr would leave a strong signature on the integrated SED of any cluster, which is clearly incompatible with our observations of W3 (cf. \S \ref{sec:discussion}).

\item For this experiment we have restricted our second star-formation episode to a FWHM = 100 Myr. If we were to increase the FWHM of the second star-formation episode, we would increase the difference between the SEDs of the synthetic clusters and that of W3, as we would be increasing the fraction of younger populations relative to the original instantaneous burst.
\end{enumerate}

The rationale behind these choices is to favour all these extended star-formation scenarios, as we are trying to minimize the differences in the SEDs between an extended star formation episode and an instantaneous burst (our best fit SFH for W3).

In G14, the authors used the results from a simulation called SG-R1 by \cite{D08} to describe the dynamical evolution of \iac .\footnote{This simulation most likely is inadequate to describe \iac\ in the SMC/LMC as we discuss in \S \ref{sec:sim}.} This particular simulation was chosen due to the ``agreement" in the reproduction of the mass fractions formed during the second, extended and first, instantaneous star-formation episodes inferred by G14 from the \mbox{eMSTOs} of \iac\ at their current ages (i.e. mass ratio between the second and first generation stars $\mbox{M}_2:\mbox{M}_1 \approx$ 2:1 observed today, 1--2 Gyr after their birth). For our synthetic clusters we take the values for $\mbox{M}_2:\mbox{M}_1$ from this simulation at younger ages, i.e. between $100 \lesssim t/\mbox{Myr} \lesssim 1000$. These range from $\mbox{M}_2:\mbox{M}_1 =$ 0.3:0.7 (at $\sim100$ Myr) to $\mbox{M}_2:\mbox{M}_1 =$ 0.5:0.5 (at $\sim1$ Gyr).

\begin{figure}
\includegraphics[width= 84mm]{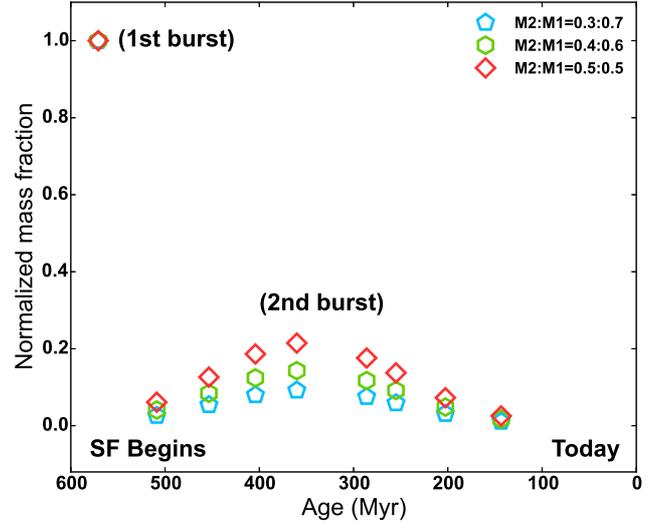}
\caption{
Star formation histories of three synthetic clusters. The pentagons, hexagons and diamonds represent the SFH of composite stellar population (CSP) clusters with mass ratios of $\mbox{M}_2:\mbox{M}_1=$ 0.3:0.7, 0.4:0.6 and 0.5:0.5 respectively. For all CSP clusters the first star formation episode is an SSP of 570 Myr. The second episode of star formation follows a Gaussian distribution centred at 350 Myr with FWHM = 100 Myr.}
\label{sfh}
\end{figure}

\begin{figure*}
\includegraphics[width= 180mm,height=84mm]{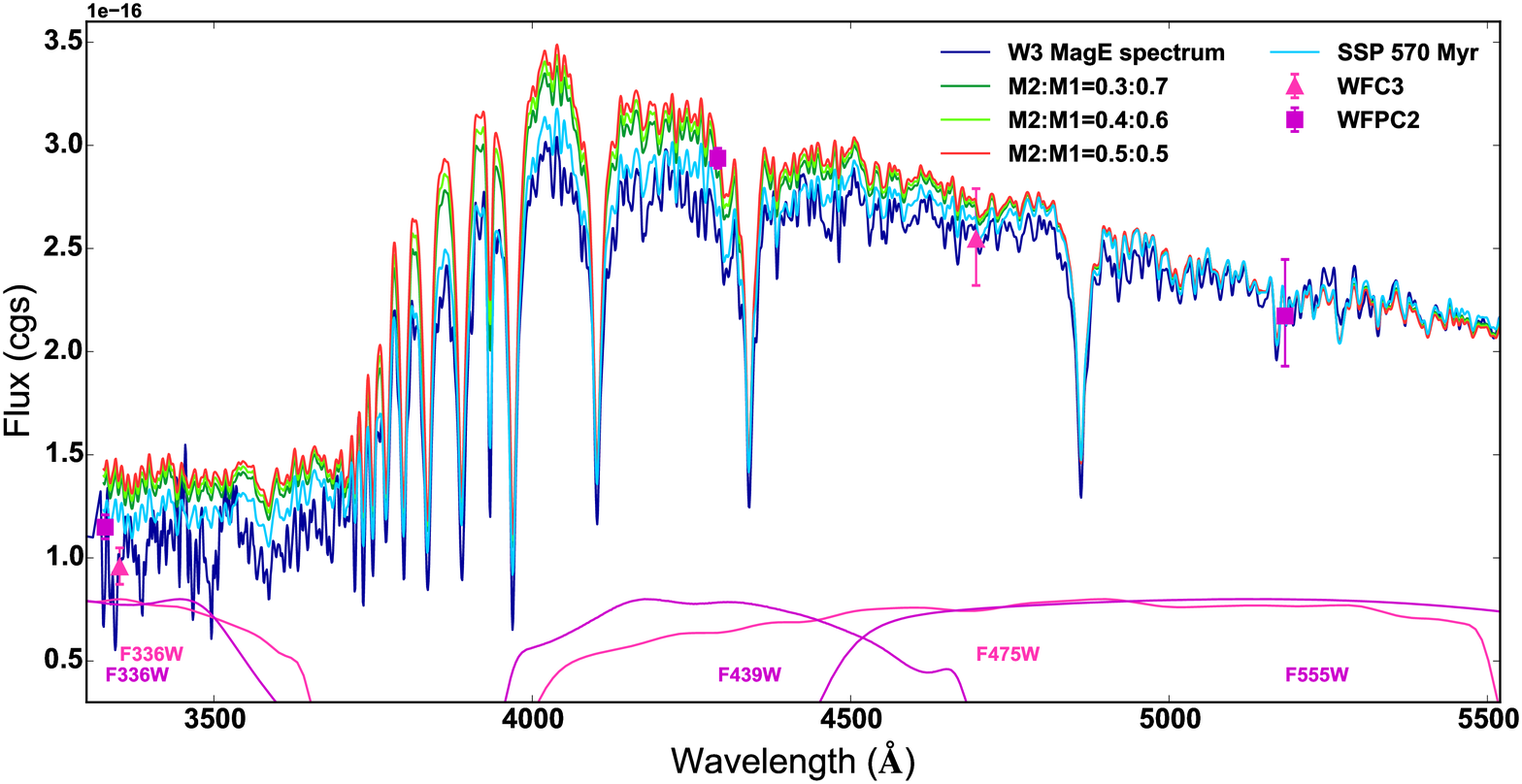}
\caption{
The dark blue line represents the integrated MagE spectrum of W3. \emph{HST} WFPC2 and WFC3 photometry of W3 is represented with squares and triangles, respectively. The model spectrum recovered by \dinbas~for W3 is shown in cyan (SSP of age 570 Myr).  Dark green, green and red lines represent the synthetic SEDs of CSP clusters with a mass ratio of $\mbox{M}_2:\mbox{M}_1=$ 0.3:0.7, 0.4:0.6 and 0.5:0.5, respectively.  The various SEDs are all normalized to the flux of W3 at the effective wavelength of the $F555W$ filter. The transmission curves of the filters used in this work (in this  wavelength range) are shown at the bottom of the figure.}
\label{SEDs1}
\end{figure*}

\begin{figure}
\includegraphics[width= 84mm]{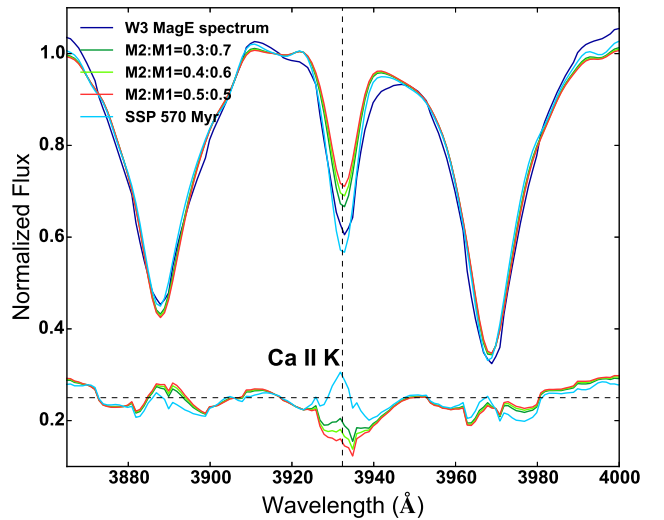}
\caption{
Same as Fig. \ref{SEDs1}, but shown as a close-up around the age-sensitive Ca {\sc ii} K line. For this figure the continuum of all SEDs had been normalized for a better comparison of the line profiles. All CSP synthetic clusters miss the depth and profile of this line. The \dinbas\ solution (570 Myr SSP) yields the best representation of this line and of the colours of W3. The bottom of the figure shows the residuals of the spectral fits.}
\label{CaIIK}
\end{figure}

Figure \ref{sfh} shows the SFHs of synthetic multiple-generation clusters for three different mass ratios.  We will refer to these synthetic clusters with extended star-formation episodes hereafter as composite stellar population (CSP) clusters.  The synthesised SEDs of these CSP clusters are shown in Figs. \ref{SEDs1} and \ref{CaIIK} together with the observed spectrum of W3.

\subsection{Experiments with \citet{Goudfrooij:2014p2918} pseudo-age distributions}
\label{pseudo-age}
Although for a massive cluster of W3's age (570 Myr) the SG-R1 simulation shows that most of the cluster mass comes from the stars of the first generation, we also explore the possibility that \emph{this entire generation} was already lost during the early dynamical evolution of this cluster, leaving just the observed age distribution derived by G14 from the eMSTO stars.

\begin{figure}
\includegraphics[width= 84mm]{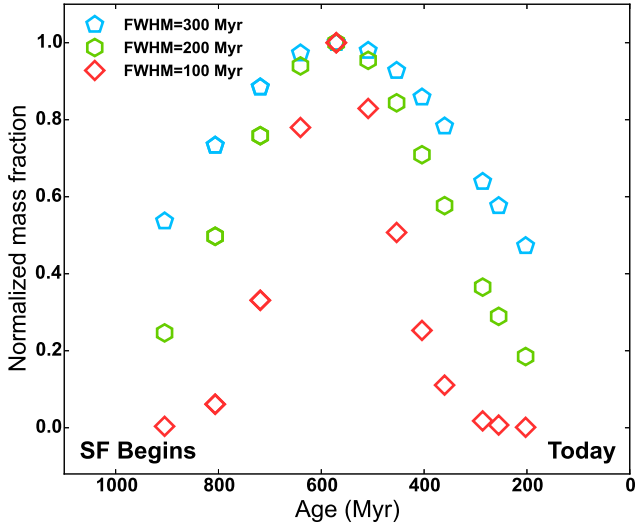}
\caption{
Pseudo-age distributions from G14 for three synthetic clusters. The pentagons, hexagons and diamonds represent the SFH of CSP clusters with FWHMs equal to 300, 200 and 100 Myr, respectively.  All distributions are centred at an age of 570 Myr.}
\label{sfh2}
\end{figure}

\begin{figure*}
\includegraphics[width= 180mm,height=84mm]{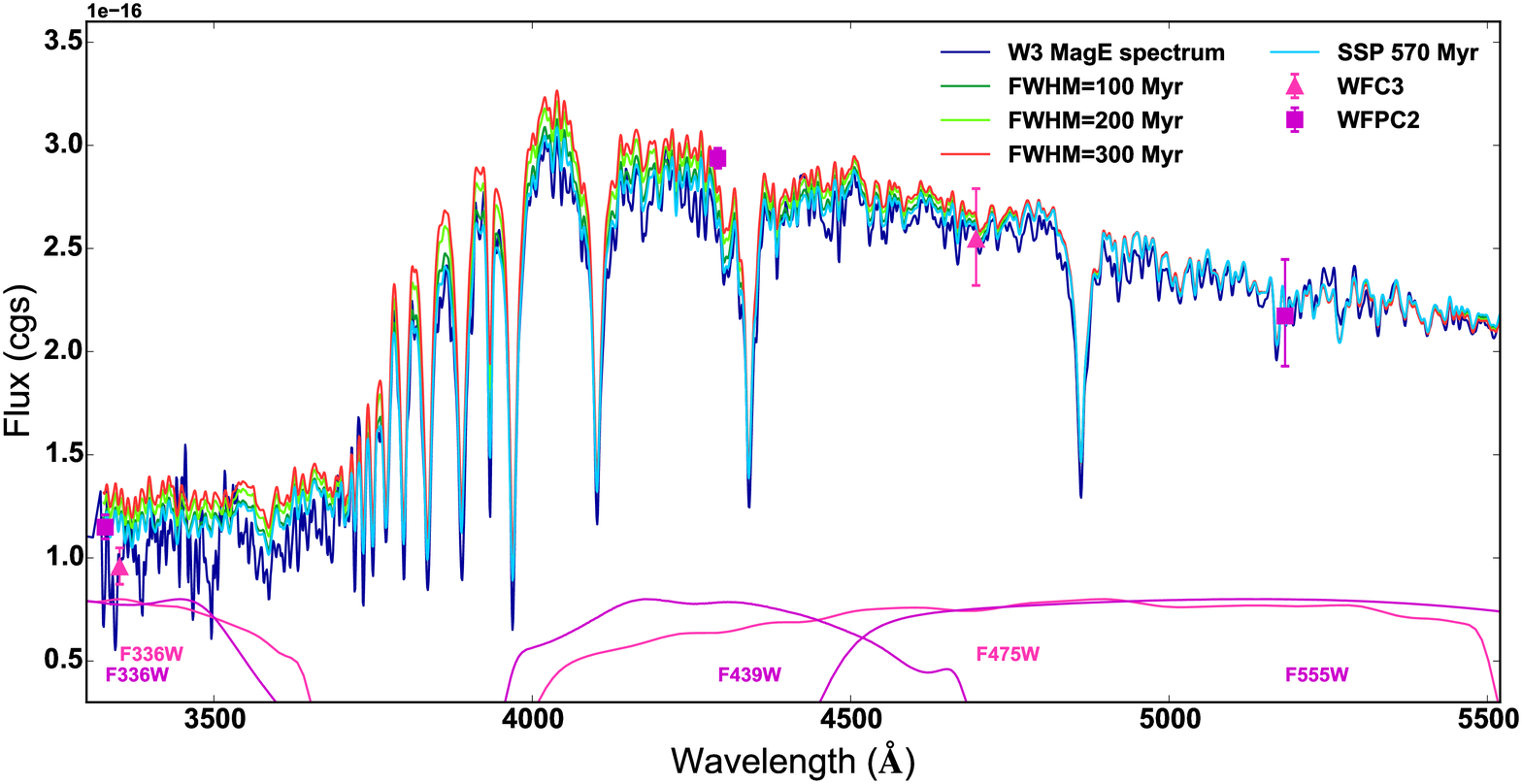}
\caption{
Similar to Fig. \ref{SEDs1}, but here we show in red, green and dark green the synthetic SEDs of CSP clusters built following the pseudo-age distributions from G14 with FWHM = 300, 200, and 100 Myr, respectively.}
\label{SEDs2}
\end{figure*}

\begin{figure}
\includegraphics[width= 84mm]{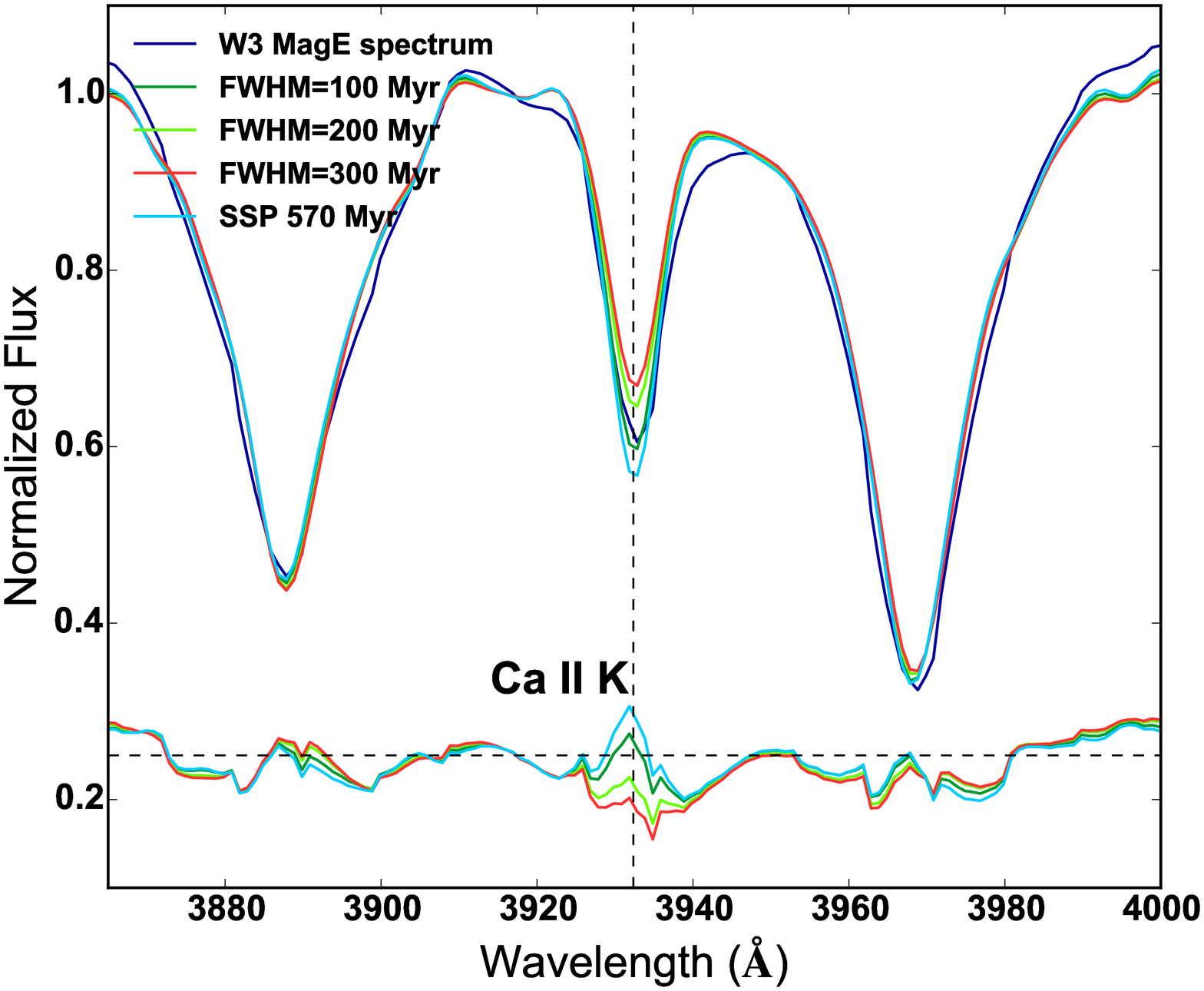}
\caption{
Same as Fig. \ref{CaIIK}, but for model CSP clusters following the pseudo-age distributions from G14.  In the same way as with the predicted SFHs from G14, the \dinbas\ solution (570 Myr SSP) and the CSP with FWHM $\le 100$ Myr yield the best representation of the Ca {\sc ii} K line of W3 when compared to the SEDs of CSP with FWHM $> 100$ Myr clusters that follow the G14 pseudo-age distributions. The bottom of the figure shows the residuals of the spectral fits.}
\label{CaIIK2}
\end{figure}

To test this possibility, we built a second set of CSP clusters, this time following the present-day age distribution (i.e. pseudo-age distributions in G11a,b and G14). For this we assumed a simple Gaussian distribution centred at 570 Myr and with FWHM ranging from 300 to 100 Myr (cf. Fig. \ref{sfh2}). The resulting SEDs are shown in Figs. \ref{SEDs2} and \ref{CaIIK2}, where they are compared directly with the observed W3 spectrum.

\section{Discussion}
\label{sec:discussion}

Figure \ref{SEDs1} shows a portion of the MagE spectrum of W3  together with overplotted synthetic spectra of the best SSP \dinbas\ solution as well as of the CSPs with SFHs shown in Fig. \ref{sfh}.  For this figure, we have normalized the flux of all our synthetic cluster spectra to the observed flux of W3 at WFPC2's $F555W$.  Obviously, the MagE spectrum agrees well with the WFPC2 and WFC3 photometry (data points with error bars).  This confirms that the wiggles in flux present in the MagE spectrum are negligible compared to the overall SED.  This figure shows that the \dinbas~solution (570 Myr SSP) is a very good representation of the observed SED of W3, while the synthetic CSP clusters fail to reproduce the optical colours of this cluster as these CSPs are too blue ($\sim10\%$ brighter for $3700< \lambda/\mbox{\AA}<4000$).  The offsets between the \dinbas\ solution and the MagE spectrum presented in Fig. \ref{SEDs1} are in part due to the wiggles in the continuum mentioned in \S \ref{sec:data}.  Once the \dinbas\ solution and the MagE data are both continuum-normalized, the spectra match very well as shown in Fig. \ref{dinbasfit}.

Besides analysing the overall shape of the SED of W3 and its agreement with the SED of synthetic SSP and CSP clusters, we can also analyse the behaviour of individual spectral features that are sensitive to age. Figure \ref{CaIIK} shows the same SEDs as displayed in Fig. \ref{SEDs1}, but now centred on a narrow region around the Ca {\sc ii} K line, which is very sensitive to age (for a given metallicity, this line gets deeper with age).  Figure \ref{CaIIK} illustrates once more how the CSP clusters fail to reproduce the observed characteristics of W3 (this time the profile of the Ca {\sc ii} K line), and although the SSP of 570 Myr does not reproduce perfectly the depth of this line (as it is a little bit deeper), it still does a better job than any of the CSPs. As already stated, this particular line gets deeper with age, whence even a small young component in a CSP would make this line appear significantly shallower. The  differences between the W3 spectrum and the best SSP solution in the profile of this line, could be due to the finite sampling in age of the SSP templates used for this work. If this were the case, the precise age of this cluster would be somewhere between the template of 570 Myr and the next youngest (508 Myr) one, as this spectrum will have about the same depth of the Balmer lines but the Ca {\sc ii} K line will be a little bit shallower than that of the 570 Myr template.
%while the SSP cluster of 570 Myr still better reproduces the observed spectrum of W3. 

We note that the \emph{extent} of our second episode of star formation (i.e. when the extended burst begins and ends) is limited to $\sim350$ Myr (see Fig. \ref{sfh}), which is about the FWHM of the extended star-formation episode proposed to explain the eMSTOs of some \iac\ according to G11a,b and G14.  If we were to consider a more extended episode of star formation (i.e. $\sim700$ Myr as suggested by G14), the differences in the SEDs and the Ca {\sc ii} K line between W3 and the synthetic CSP clusters would become more marked due to the inclusion of even younger components to the CSP.
In other words, \emph{even with these favorable assumptions, like a short second star-formation episode (FWHM=100 Myr) and the truncation of the star formation at younger ages (no star-formation in the last 100 Myr, see Fig. \ref{sfh}), we still find considerable differences between the SEDs of a cluster with a SSP and a cluster that experienced such extended star-formation event.}

We can use the experiments with the pseudo-age distributions (\S \ref{pseudo-age} and  Fig. \ref{sfh2}) to place constraints on any extended SFH that may be present in W3 (i.e. the age resolution that our data offer). Extended star formation leaves a second-order imprint on a cluster's SED due to the nonlinear evolution of colour. A cluster with an extended burst of star formation centred at lookback time $t$ will be bluer than an SSP of age $t$; the effect scales as $(\Delta t/t)^2$, where $\Delta t$ is the duration of star formation. Extended SFHs have
been inferred from eMSTOs in the CMD, where the effect is instead first-order: the width of the MSTO increases linearly with the duration of
star formation.

%Figure \ref{SEDs2} shows how these synthetic clusters compare to the observed SED of W3. 
Figure \ref{SEDs2} shows how the SEDs of synthetic clusters with the extended SFHs of Fig. \ref{sfh2} compare to the observed SED of W3.
 At first glance, one can see that the CSPs with the long star-formation episodes (FWHM = 200 and 300 Myr) are easily distinguished from the W3/SSP SED, since the flux in $3700< \lambda/\mbox{\AA}<4000$ is again about 10\% higher than that of the W3 or SSP SEDs. But this time, the CSP cluster with the narrowest star formation episode (i.e. FWHM=100 Myr) is not very different from the SSP or W3's SED.

We conclude the same, if instead we focus on the Ca {\sc ii} K line, as shown in Fig. \ref{CaIIK2}, we can see that the CSPs with FWHM $>100$ Myr never reproduce the profile of this line better than the single SSP spectrum.  We conclude from this that in order to produce a CSP that has an SED and shape of the Ca {\sc ii} K line similar to what is observed in W3,
 we would need to (1) decrease the contribution of the main (first) burst proposed by G14, either by reducing its mass or by reducing the
  period between the first and second burst to be short enough so that both bursts would
 be indistinguishable, and (2) reduce the extent of the second star-formation period, down to a limit where it approaches 
 the SED of an SSP.  Figure \ref{CaIIK2} illustrates that this would be the case for a CSP following a Gaussian distribution with a FWHM $\le100$ Myr.\footnote{This upper limit for the extent of any prolonged star formation episode within W3 is consistent with the uncertainties/degeneracies in age estimated for this cluster cf. \S \ref{sec:dinbas} and Appendix \ref{sec:deg}.}

We note that the mean FWHM age spread claimed by G14 for a sample of 18 \iac\ is 375 Myr and the shortest FWHM is 200 Myr.  These values are well above our derived limit where one could start confusing the SED of an SSP and of a brief ($\le100$ Myr) CSP event.  
Stellar rotation can mimic a fractional age spread of $\sim$20\%, or $\sim$100 Myr at the age of W3; Fig. \ref{SEDs2} shows that W3 is perfectly consistent with such a pseudo-age spread.  It is, however, strongly inconsistent with the $\sim$300--400 Myr durations inferred from the eMSTOs in older LMC clusters.

%In other words,\emph{ the only way a CSP's SED can mimic the one of an SSP for this particular cluster is a Gaussian burst, centred at the \dinbas\ solution (570 Myr) with a FWHM $\le 100$ Myr.  This ``permitted" age spread is significantly smaller than the spreads proposed by G14 to explain the origin of the eMSTOs in \iac\ of the Magellanic Clouds.}

\subsection{Is the SG-R1 simulation suited to describe eMSTO clusters?}
\label{sec:sim}

The simulation chosen by G14 to describe the dynamical evolution of \iac\ in the SMC/LMC displaying eMSTOs may not reproduce accurately their mass losses. This simulation, described in \cite{D08} and named SG-R1, was originally developed to study the dynamical evolution of $\sim10$ Gyr old Galactic globular clusters in a strong tidal field.  Due to this, there are some issues that may make it unsuitable for
studies of intermediate-age SMC/LMC clusters, such as: 

\begin{itemize}

\item The second-generation stars in this simulation have only masses between 0.1 and 0.8 \msun.  Yet, the estimated masses of stars that populate the turnoff of intermediate-age (1--2 Gyr) clusters are all $>1$ \msun. \emph{Hence, this simulation would not feature any stars visible today in the regions of colour--magnitude diagrams where the eMSTO is observed in \iac.}

\item The main reason that led G14 to choose this particular simulation for describing eMSTO clusters is that the mass ratio between second- and first-generation stars yielded at the age of \iac\ is very similar to the ratio inferred from the SFH derived from the eMSTO.  This ratio is $\mbox{M}_2:\mbox{M}_1 \approx$ 2\,:\,1 for present-day eMSTO clusters.  \emph{However, Figs. 15 and 16 of \cite{D08} show that one only gets this ratio for stars in the mass range $0.1\le \mbox{M/M}_\odot \le 0.8$}.  If one takes into account that the stars of the first generation in this simulation cover masses between 0.1 and 100 \msun\ following a standard (i.e. Kroupa) IMF, the actual mass ratio between the second- and first-generation stars decreases as there are more stars from the first generation than just the ones between $0.1\le \mbox{M/M}_\odot \le 0.8$.

\item The cluster in the simulation SG-R1 is tidally limited, i.e. the stars in the cluster are distributed up to the radius where a star is equally bound to the cluster and the Galaxy.
 This setup provides an extremely efficient way to lose stars. However, \emph{all the \iac\ studied in G14 have tidal radii between 4.9 and 43 times larger than their core (and effective) radii according to these authors.} This fact makes mass loss in these clusters very inefficient.

To put this in context, we emphasize that the scenario of G14 requires retaining the first generation of stars (most of the cluster mass) at younger ages in order to provide enough gravitational potential to hold on to the gas expelled from evolved stars and to accrete gas from their surroundings. This gas will eventually fuel the extended (second) star-formation episode responsible for the eMSTO. Figures 2, 3 and 4 of G14 show that for these 1--2 Gyr old clusters, this second episode stopped forming stars 800--1500 Myr ago (depending on the cluster). Hence, the first-generation stars can only be lost after these ages, when they are no longer needed to hold on to the material to form the second burst. In other words, \emph{this implies a strong change of the tidal potential hosting each of these clusters in the last 800--1500 Myr, for which---to the best of our knowledge---there is no evidence.} 

\item Another critical factor is that the simulation SG-R1 assumes that the cluster is sitting in the Galactic potential tidal field, at a galactocentric distance of 4 kpc.  Yet, the tidal fields of the SMC and LMC are significantly weaker than that of the Galaxy.  Hence, the disruption derived from this simulation will be significantly overestimated compared to the actual one suffered by the eMSTO clusters in the SMC/LMC. We figure that this simulation could only apply to the eMSTO clusters if their densities at
young ages were about two orders of magnitude lower than the densities of YMCs observed today (see Appendix \ref{sec:disruption}).

\end{itemize}

We conclude that the simulation SG-R1 by \cite{D08} is unsuitable for describing the dynamical evolution of SMC/LMC clusters with an eMSTO.

\subsection{NGC 1856: a YMC with an eMSTO}
\label{sec:1856}

The young cluster NGC 1856 represents a very interesting peace of the puzzle of the origin of the eMSTO. As mentioned before, this massive ($\sim10^5$ \msun) young cluster with an age of $\sim300$ Myr is significantly younger than any of the clusters previously reported with eMSTOs (with ages usually between 1--2 Gyr). The young age of this cluster allows us an unprecedented opportunity to understand the early evolution of a cluster with an eMSTO. In this section we combine the information derived by \cite{Milone15,Correnti15} from the eMSTO of NGC 1856 with previous studies of YMCs in order to place some constrains on the origin of the eMSTO.

In a previous study we analysed the YMC NGC 34:\,S1 \citep{cz14}. Using \dinbas\ we concluded that the SFH of this cluster is consistent with an SSP of age $100 \pm 30$ Myr and mass $1.9\pm 0.2 \times 10^7$ \msun.  We were able to rule out any significant episode of star formation in the last 70 Myr of the cluster. G14 claim this is consistent with their scenario, as perhaps 100 Myr
%Regarding this result, G14 suggest that perhaps 100 Myr
 may not be enough for a second episode of star formation to take place, as the Lyman-Werner photons from the first-generation stars prevent the gas from cooling down to form stars \citep{cs11}. 

If the suggestion of G14 indeed applies, then a break in the star formation lasting $\sim100$ Myr would not be consistent with the age spreads inferred from the eMSTO of the young cluster NGC 1856.
%If this was the case, the result from NGC 34:\,S1 would not be consistent with the age spreads inferred from the eMSTO of the young cluster NGC 1856.
  For the age of this cluster ($\sim300$ Myr) a $\sim100$ Myr delay between the first- and second-generation stars would be readily observable in its CMD.  However, the distribution of stars in the eMSTO of this cluster, and the inferred age distributions, seem to be continuous with no apparent gaps \citep{Milone15,Correnti15}. This could mean that the beginning of the extended star-formation episode responsible for the eMSTO varies from cluster to cluster in a very peculiar way, as it has never been observed to be ongoing in \emph{any} YMC.  That is, no evidence of ongoing star formation has been found in a study of $\sim130$ young (10--1000 Myr) massive ($10^4-10^8$ \msun) clusters \citep{Bastian:2013p2022}.   Nor has any evidence been found of gas reservoirs within YMCs that could fuel extended star-formation episodes with the masses suggested by these scenarios (e.g. \citealt{Bastian14,cz15}). Hence, the alternative explanation is that it cannot be an age spread that is responsible for the eMSTO.

\cite{Niederhofer15b} found a correlation between the width of the eMSTO (or inferred age spread) and the age of the clusters in their sample, suggesting an evolutionary effect. For instance, a young cluster like NGC 1856 has an inferred MSTO age spread of 140 Myr (c.f. \citealt{Milone15}), while an older cluster (1 Gyr) like NGC 2108 has an inferred age spread of 230 Myr spread and an even older (1.45 Gyr) cluster like NGC 411 shows an MSTO spread equivalent to 516 Myr (c.f. G14).

Evolutionary effects, like stellar rotation, have been proposed as an alternative to age spreads for the origin of the eMSTO (c.f. \citealt{Bastian09,Yang13,Li14}). \cite{Niederhofer15c} showed that stellar models that include rotation can reproduce an evolution of the MSTO morphology in time. Furthermore, they also show that if these evolutionary effects (rotation) were to be interpreted as age spreads, the inferred age spreads from the eMSTO are in agreement with the ones claimed by the authors supporting the age spread scenario e.g. G14.

%\textbf{The relatively small age spread ($\sim140$ Myr) inferred from the MSTO of young clusters like NGC 1856 compared to larger ones ($\sim 375$ Myr) inferred from intermediate-age clusters, makes it consistent with the prediction of stellar models that account for rotation.}

Supporting this interpretation is the recent analysis of the CMD of NGC 1856 by \cite{Dantona15}, which suggests that the complex MSTO of this cluster is due to two populations of the same age ($\sim$ 350 Myr), one composed mainly of very rapidly rotating stars ($\omega=0.9 \omega_{\mbox{crit}}$), while the other is composed of slowly/non-rotating stars. %However, the predicted red clump of this cluster according to the solution inferred from the eMSTO (i.e. two populations with different rotational velocities) does not reproduced very well the observations, as the observed CMD seems to be missing the distinct signature of the slowly/non-rotating population. Having said that, we note that the reproduction of this evolutionary phase (red clump) encounters some challenges in the stellar evolution modes as discussed by D?Antona et al. (2015).}

In this context, we can use W3 to test the age spread or the rotation scenarios.  As seen in the previous sections, the maximum age spread present in W3 is at most 100 Myr.  In the age spread scenario suggested by G14, for a cluster with such a high escape velocity, the expected age spread is typically $\sim375$ Myr, inconsistent with the observations.  On the other hand, in the rotational scenario, the expected MSTO spread would be the equivalent of $\sim 150-200$ Myr.  However, the post-main sequence features, which contributed significantly to the integrated light are expected to have age spreads equivalent to $<100$ Myr \citep{Niederhofer15c}.  Hence, our observations are inconsistent with the age spread scenario, but consistent with those expected from the rotational scenario.

\section{Summary and Conclusions}
\label{sec:conclusions}

We have used the SED of W3, a YMC in the merger remnant NGC 7252, and have compared it with the SED of synthetic clusters constructed with the SFHs that are proposed to explain the eMSTOs of \iac.  We find that the SED of this cluster is consistent with that of an SSP of age $570^{+70}_{-62}$ Myr, mass $1.13^{+0.14}_{-0.13}\times10^8$ \msun\ and current central escape velocity above 193 km s$^{-1}$.

A key argument of some of the scenarios that have been proposed to explain the eMSTOs of \iac\ is based on the fact that these clusters all have masses above $10^4$ \msun.  These scenarios also assume that the eMSTO clusters were a factor of 10--20 more massive at birth and had escape velocities $>15$ km s$^{-1}$, enabling them to retain the gas that fueled the extended episode of star formation responsible for the observed eMSTO (cf. \citealt{Keller11}).  However, our results are in strong contradiction to the prediction of these scenarios,
%appear to disagree with this hypothesis, 
given that the SED of W3 does not match the model SED of young clusters with an age spread similar to any of those suggested for eMSTO clusters. %, in strong contradiction to the prediction of these scenarios. 
Yet, W3 is the most massive young cluster known to date, and its mass ($\sim10^8$ \msun) and escape velocity ($>193$ km s$^{-1}$) exceed by orders of magnitude the masses/escape velocities (and the expected birth masses/escape velocities) of \iac\ showing eMSTOs.  The near lack of significant extinction in this cluster ($A_V=0.083$) is also in conflict with the properties of a young cluster hosting a massive reservoir of cool gas \citep{Longmore15}, suggesting that currently there is no cold gas that could fuel an extended star-formation episode in the near future.

Of course, there is a possibility that this specific cluster has some intrinsic property or that there is something peculiar in its environment that has prevented any extended episodes of star formation to take place.  Yet, note that these results are in perfect agreement with previous studies of YMCs, as they all point towards a SFH of a single burst with a negligible extent.  Also, in the scenarios investigated here there is no explicit requirement regarding
% mention nothing explicitly about 
 the eMSTO cluster environments that would make the results we have obtained from W3 irrelevant for constraining the origin of this phenomenon.

Additionally, the open clusters Hyades and Praesepe---composed of $\sim\,$300 and $\sim\,$1000 members, respectively, and each $\sim\,$800 Myr old \citep{Perryman98,Kraus07}---are known to have MSTOs
%turnoffs
 that are inconsistent with a single isochrone, which has been attributed to spreads of a few hundred Myr in age \citep{Eggen:1998p2928}. \citet{Brandt15a} have recently proposed that age spreads are not needed to explain the broadened MSTOs in these clusters, since stellar models that include \emph{rotation} can reproduce the same morphology with a single generation of stars. 
 This mechanism has also been explored in the past to explain the origin of eMSTO clusters in the SMC/LMC (e.g. \citealt{Bastian09}). 

However, given the relatively small number of stars in the Hyades and Praesepe, it is hard to tell if the broadening of the turnoff observed in these open clusters is the same phenomenon as that observed in the significantly more massive ($ > \sim10^4$ \msun) \iac\ displaying an eMSTO.  If this were the case, this would also challenge the suggestion
 that the cluster's gravitational potential well retained the gas long enough to have these extended star formation episodes, as it is highly unlikely that these open clusters had the necessary elevated escape velocity at birth. 

% \textbf{Our results from the analysis of the SED of W3 are consistent with the interpretation of the eSMTO of \iac\ as the effects of stellar rotation.}
  
\cite{Piatti15} have recently found evidence for eMSTO in extremely low mass clusters (< 5000 \msun) in the LMC, which shows that cluster mass cannot be the essential parameter. Similar hints of eMSTOs can been seen in CMD of the low-mass
Galactic open clusters NGC 752 and Tombaugh 1
% Similar evidence have also been observed recently in open clusters NGC 752 and Tombaugh 1 c.f.
  \citep{Twarog15,SalesSilva15}.

%\textbf{Our results from the analysis of the SED of W3 are not consistent with age spreads as the origin of the eMSTO in \iac\, but consistent with the effect of stellar rotation as the origin of the eMSTO.}

Finally, we would like to reiterate the important point made by G11 that, if
%Finally, G11a correctly point out that, if
 the eMSTOs observed in \iac\ are due to the same mechanisms proposed to explain the chemical abundance patterns in GCs, then one would expect to see the same light-element abundance variations in \iac.  However, different abundance analyses have shown no signs of these abundance patterns in the LMC \iac\ NGC 1806, 1651, 1783, 1978, and 2173 \citep{Mucciarelli:2008p2339,Mucciarelli:2014p2575}.  Some of these clusters show an eMSTO (e.g. NGC 1783, 1651 and 1806) and have stellar masses comparable to GCs (all of these \iac\ are above $10^5$ \msun).  All of these considerations suggest, again, that the eMSTO phenomenon is unrelated to the chemical anomalies found in old GCs.

\section{Acknowledgements}

We would like to thank P. Goudfrooij, C. Li and R. de Grijs for enlightening discussions and comments on an earlier version of this manuscript. This paper includes data obtained with the Clay 6.5-m Magellan Telescope located at Las Campanas Observatory, Chile.
NB is partially funded by a Royal Society University Research Fellowship and an European Research Council Consolidator Grant (Multi-Pop - 646928). FS gratefully acknowledges support from the Carnegie Institution for Science. 
GB acknowledges support for this work from the National Autonomous University of M\'exico (UNAM), through grant PAPIIT IG100115. JMDK is funded by a Gliese Fellowship. This work was performed in part under contract with the Jet Propulsion Laboratory (JPL)
funded by NASA through the Sagan Fellowship Program executed by the NASA Exoplanet Science Institute.

\bibliographystyle{mnras}
\bibliography{cz16a}

\appendix

\section{Degeneracies of the continuum normalized fit}
\label{sec:deg}

As mentioned in \S \ref{sec:dinbas}, we use the SSP solution from the \dinbas\ fit to the continuum normalized spectrum, as the best estimate of the age of NGC 7252: W3. This fit yielded an age for this cluster of 570 Myr. In this appendix we look for possible degeneracies in our age estimate (i.e. if some other combinations of multiple populations reproduce the W3 spectrum equally well as our best solution), conducting fits to our data using grids of synthetic multiple-population clusters.

This experiment is the same we carried out to study the degeneracies on the SFH of NGC 34: S1 in \cite{cz14}. Here we built a grid of synthetic cluster spectra, in which each element represents a cluster with two star formation events. These spectra were built using the same continuum normalized \cite{bc03} models we used for our DynBaS fits (cf. \S \ref{sec:dinbas}). In this grid each synthetic cluster consists of a massive population with always the same age, 570 Myr (Pop. I from here on), followed or preceded by a less massive second population of a different age (Pop. II). The ages for Pop. II in this grid range from 1 Myr to 1 Gyr and are distributed almost uniformly in log space. While the masses of Pop. II could take values ranging from 10 to 90 per cent of the mass of Pop. I.

\begin{figure}
\includegraphics[width= 84mm]{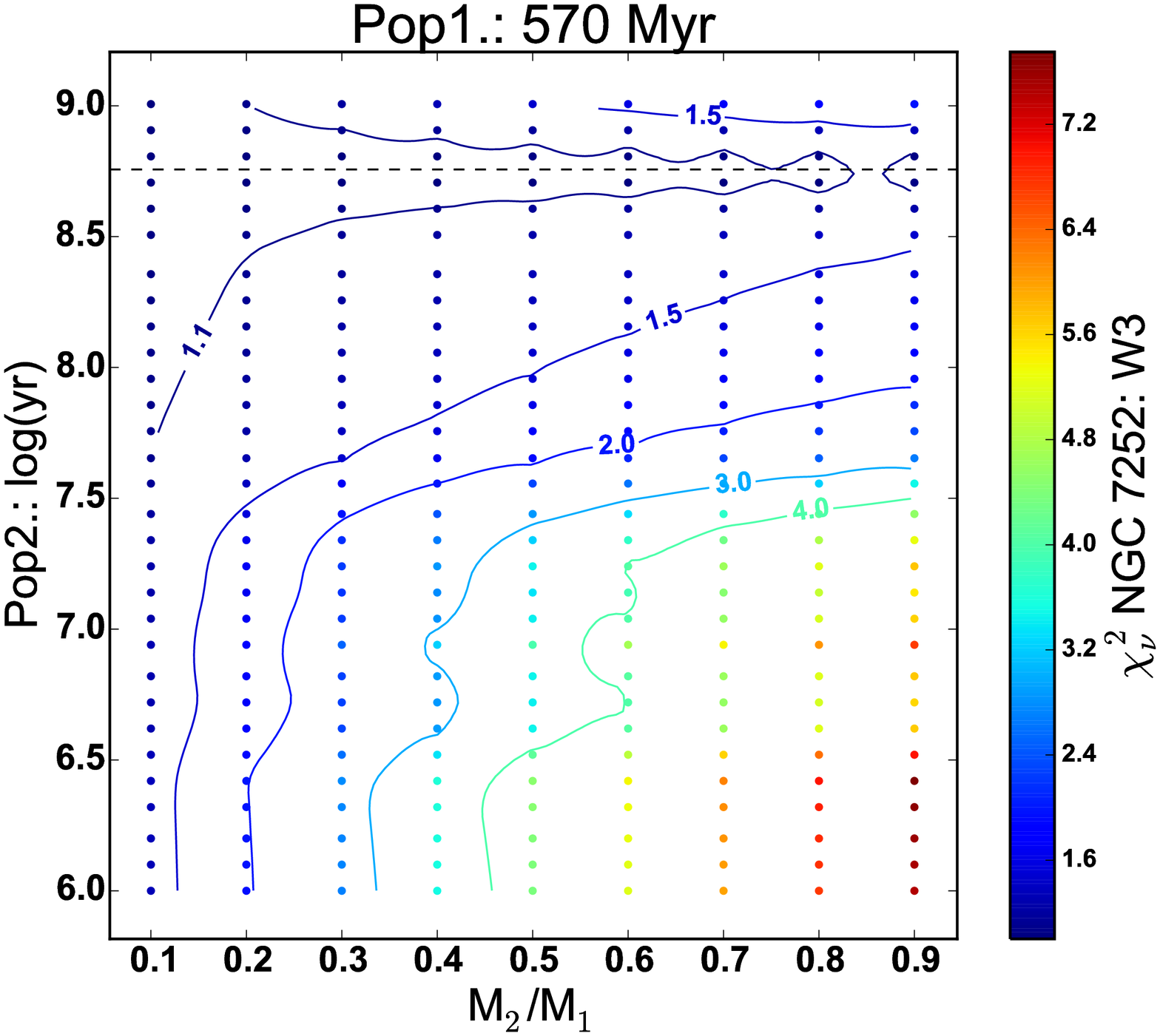}
\caption{
Results of fitting the normalized spectrum of W3 with each element of a grid of synthetic cluster spectra. The Pop. I age for each element is always the same 570 Myr (i.e. our best \dinbas\ SSP solution), which is represented with the dashed horizontal line in this plane. The vertical axis represents the age of the secondary (less massive) population, Pop. II, while the horizontal axis denotes the mass ratio between the first and second population. In colour we represent the $\chi_\nu^2$ for each of these fits. For $\chi_\nu^2>1.1$, we can spot the differences by eye between the synthetic clusters and the data (specially in the Ca{\sc ii} K line), so those solutions are immediately excluded.}
\label{grid}
\end{figure}

In Fig. \ref{grid} we show the results of the fits of the continuum normalized MagE spectrum of W3 to each of the elements in this grid. In this figure we colour coded the solutions as a function of their $\chi_\nu^2$. The contours denote constant values of $\chi_\nu^2$. For practical reasons, we have normalized all these values dividing each of them by the $\chi_\nu^2$ of our \dinbasuno\ solution (i.e. SSP 570 Myr). We found that for fits with $\chi_\nu^2>1.1$, it is possible to distinguish by eye that the spectral fits are poor (i.e. fail to reproduce the depths/profiles of some Balmer lines and the Ca{\sc ii} K line), and such solutions are excluded. For reference, the differences between solutions with $\chi_\nu^2<1.1$ and $\chi_\nu^2> 1.1$, are similar (or greater) to the differences between the FWHM = 100 and 200 Myr spectra and the spectrum of W3 in Fig. \ref{CaIIK2}, respectively.

Overall, this parameter space has a similar behaviour as the one found for NGC 34: S1 in \cite{cz14}. In the sense that we do not see any other region with a local $\chi_\nu^2$ minimum which could host/hide another solution that represents a fit as good (or better) than the \dinbas\ SSP solution. This figure also shows that the region with $\chi_\nu^2<1.1$ (i.e. the region where it is not possible to distinguish a multiple population solution from a SSP solution) basically comprises all the synthetic clusters with a secondary burst (Pop. II) of age 508 and 640 Myr (which are the row of dots bellow and above the dashed line respectively) regardless of their mass. This region lies exactly within the uncertainties reported in \S \ref{sec:dinbas} \dinbas\ fit solution, $570^{+70}_{-62}$ Myr. This is also in agreement with the upper limit of 100 Myr for the width of an extended star formation event centred at 570 Myr, where this mulitple population solution represent a fit as good as the SSP solution (cf. Fig. \ref{CaIIK2} and the discussion in the text about this figure).

From this experiment, we conclude that the possible degeneracies in the SFH of this cluster, i.e. any multiple population solution as good as our SSP solution, lie within the uncertainties in the age of W3 reported in \S \ref{sec:dinbas}. We note that these uncertainties/degeneracies are significantly smaller than the age spread expected for this cluster according to the scenarios that attribute the eMSTO in \iac\ to extended star formation events.

\section{Cluster disruption in the SMC/LMC}
\label{sec:disruption}

Here we discuss the feasibility of applying the strong Galactic tidal field of the \citet{D08} models to the intermediate-mass clusters in the SMC/LMC studied by G14.

Most of the dynamical mass loss in the \citet{D08} model of multiple-generation clusters is induced by stellar evolution, i.e. by the change of the gravitational potential due to massive stars ending their lives. This mass loss depends sensitively on how extended a cluster is --- \citet{D08} achieve the high second-to-first generation ratio required by G14 only by assuming that the cluster fills its Roche lobe (their SG-R1 simulation), allowing it to lose stars efficiently as the gravitational potential changes. Hence, this model can only be applied to the SMC/LMC clusters if the ratio between the cluster radius and the tidal radius $r_{\rm h}/r_{\rm t}$ is the same as in the \citet{D08} model. This ratio is $r_{\rm h}/r_{\rm t}=\{0.187,0.115\}$ in Roche lobe-filling clusters with King parameters $W_0=\{5,7\}$ \citep[e.g.][]{Lamers10}. Independently of the cluster's mass, this translates directly to a ratio between the half-mass density $\rho_{\rm h}=3M/8\pi r_{\rm h}^3$ and the tidal density $\rho_{\rm t}=3\Omega^2/2\pi G$ (where $\Omega=V/R$ is the orbital angular velocity within the cluster's host galaxy, assumed to have a flat rotation curve) of $\eta\equiv\rho_{\rm h}/\rho_{\rm t}=\{153,658\}$ for $W_0=\{5,7\}$. In other words, the simulation SG-R1 {\it requires} that
\begin{equation}
\label{eq:density}
\rho_{\rm h}=\frac{3\eta\Omega^2}{2\pi G} .
\end{equation}
This density can be evaluated using the known rotation curves for the SMC and LMC \citep{Alves:2000p2938,Stanimirovic:2004p2940}.  Table~\ref{tab:density} lists $\Omega$ and $\rho_{\rm h}$ for a number of galactic environments, such as the Milky Way at a radius of $R=4~{\rm kpc}$ (as in \citealt{D08}, assuming a circular velocity of $V=220~{\rm km}~{\rm s}^{-1}$), the SMC at $R=3.5~{\rm kpc}$, and the LMC at $R=\{2,3,4\}~{\rm kpc}$. The table shows that the SMC and LMC require densities $5<\rho/{\rm M}_\odot~{\rm pc}^{-3}<70$, with median values of $\rho_{\rm h}=\{8.3,36\}~{\rm M}_\odot~{\rm pc}^{-3}$ for $W_0=\{5,7\}$, which is up to an order of magnitude lower than in the \citet{D08} model used by G14.
%\begin{table}
%\centering
% \begin{minipage}{80mm}
% \caption{Half-mass densities required for model SG-R1 in \citet{D08}.}\label{tab:density}
% \begin{tabular}{@{}l c c c c@{}}
% \hline 
% Galaxy & $R$ & $\Omega$ & $\rho_{\rm h}(W_0=5)$ & $\rho_{\rm h}(W_0=7)$ \\
%\hline
%MW & $4$ & $5.6$ & $51$ & $221$ \\
%SMC & $3.5$ & $1.8$ & $5.0$ & $21$ \\
%LMC & $2$ & $3.1$ & $15$ & $66$ \\
%LMC & $3$ & $2.2$ & $8.0$ & $34$ \\
%LMC & $4$ & $1.8$ & $5.2$ & $22$ \\
%\hline
%\end{tabular}
%Radii are in kpc, angular velocities in (100~Myr)$^{-1}$, and densities in M$_\odot$~pc$^{-3}$.
%\end{minipage}
%\end{table}
\begin{table}
% \begin{minipage}{80mm}
 \caption{Cluster galactrocentric distances, orbital angular velocities and half-mass densities required for the SG-R1 simulation in \citet{D08}.}\label{tab:density}
\begin{center}
 \begin{tabular}{@{}l c c c c@{}}
 \hline 
 Galaxy & $R$ & $\Omega$ & $\rho_{\rm h}(W_0=5)$ & $\rho_{\rm h}(W_0=7)$ \\
% & (kpc) & (100 Myr)$^{-1}$ & (M$_\odot$~pc$^{-3}$) & (M$_\odot$~pc$^{-3})$ \\
\hline
MW & $4$ & $5.6$ & $51$ & $221$ \\
SMC & $3.5$ & $1.8$ & $5.0$ & $21$ \\
LMC & $2$ & $3.1$ & $15$ & $66$ \\
LMC & $3$ & $2.2$ & $8.0$ & $34$ \\
LMC & $4$ & $1.8$ & $5.2$ & $22$ \\
\hline
\end{tabular}
\end{center}
%$^{(a)}$ assuming a circular velocity of $V=220~{\rm km}~{\rm s}^{-1}$, as in SG-R1. 
%$^{(b)}$ i.e. the original SG-R1 simulation.
Radii are in kpc, angular velocities in (100~Myr)$^{-1}$, and densities in M$_\odot$~pc$^{-3}$.
%\end{minipage}
\end{table}

To determine whether the above low densities are still reasonable for young stellar clusters, we compare them to the densities of local-Universe YMCs listed in \citet{PZ10}. The median density of the YMCs (ages $<10~{\rm Myr}$) in their comprehensive sample is $\rho_{\rm h}=10^3~{\rm M}_\odot~{\rm pc}^{-3}$, i.e.~between 1 and 2 orders of magnitude higher than the low densities required by G14 to apply the \citet{D08} model to SMC/LMC clusters.  Out of the 30 YMCs, only 3 ($\chi$Per, NGC 4038:\,W99-16, and NGC 4449:\,N-2) have densities lower than $50~{\rm M}_\odot~{\rm pc}^{-3}$. For reference, the only YMC younger than 10~Myr in the LMC (R136) has $\rho_{\rm h}>600~{\rm M}_\odot~{\rm pc}^{-3}$.

Therefore, the observed densities of recently-formed YMCs are too high for the \citet{D08} model to apply. The only way in which this application could be appropriate is if the intermediate-age clusters of G14 represent the low-density end of some larger, initial cluster population, of which the high-density YMCs were subsequently destroyed. However, there is no known mechanism that destroys high-density clusters more easily than low-density ones.  The main disruption agents in gas-rich galaxies are tidal evaporation and tidal shocks by giant molecular clouds, both of which \emph{favour} the survival of high-mass, high-density clusters \citep[e.g.][]{Kruijssen15}.

We conclude that applying the simulation SG-R1 by \citet{D08} to YMCs in the SMC and LMC requires either extremely rare or carefully-tuned conditions, making it highly unlikely that this model applies to the clusters considered by G14.

\bsp	% typesetting comment
\label{lastpage}
\end{document}